\documentclass{caosp309}

\usepackage{graphicx}
\usepackage{amsmath}

\usepackage{natbib}
\articleNo{xxx}
\pubyear{2025}
\volume{55}
\volnumber{3}
\firstpage{217}
\received{November 17, 2024}
\accepted{January 27, 2025}

\def\BibTeX{{\rm B\kern-.05em{\sc i\kern-.025em b}\kern-.08em
    T\kern-.1667em\lower.7ex\hbox{E}\kern-.125emX}}
\newcommand{\oversim}[2]{\protect{\mbox{\lower0.5ex\vbox{%
   \baselineskip=0pt\lineskip=0.2ex
   \ialign{$\mathsurround=0pt #1\hfil##\hfil$\crcr#2\crcr\sim\crcr}}}}} 
\newcommand{\simgreat}{\mbox{$\,\mathrel{\mathpalette\oversim>}\,$}} 
\newcommand{\simless} {\mbox{$\,\mathrel{\mathpalette\oversim<}\,$}} 
\begin{document}

%
\hauthor{P.\,Kroupa}

\title{Are binary-star populations regionally different?}
\subtitle{-- in memory of Sverre Aarseth --}


%
%
\author{
        Pavel Kroupa\inst{1,2}\orcid{0000-0002-7301-3377}
       }

%
\institute{
           Helmholtz-Institut fuer Strahlen und Kernphysik, University
           of Bonn,  \email{pkroupa@uni-bonn.de},
Nussallee 14-16, 53115 Bonn, Germany
         \and 
           Charles University in Prague,  \email{pavel.kroupa@mff.cuni.cz},
Faculty of Mathematics and Physics,
Astronomical Institute, V  Hole\v{s}ovi\v{c}k\'ach 2\\
CZ-18000 Praha, Czech Republic
          }

\date{March 8, 2003}

\maketitle

\begin{abstract}
  For synthesising star clusters and whole galaxies, stellar
  populations need to be modelled by a set of four functions that
  define their initial distribution of stellar masses and of the
  orbital properties of their binary-star populations.  The initial
  binaries are dynamically processed in different embedded clusters
  explaining differences in the observed populations. The approach
  summarised here, for which the Aarseth Nbody codes have been
    instrumental, allows inference of the initial conditions of the
  globular cluster $\omega$~Cen and the quantification of the stellar
  merger rate as a function of stellar spectral type, of the role of
  multiples and mergers for the Cepheid population, and predictions of
  extragalactic observables. The observability of the four initial
  distribution functions and their physical and philosophical meaning
  are also briefly raised. Evidence for the variation of these
    functions on the physical conditions of star formation and future
    steps towards extensions to include higher-order multiple systems
    are touched upon.

\keywords{stellar populations -- initial binary star population --
  $\omega$~Cen -- Cepheids -- stellar mergers}
\end{abstract}

%
\section{Introduction}

The primary distribution function defining the appearance as well as
the dynamical and elemental-abundance evolution of a stellar
population is the stellar initial mass function (IMF, $\xi(m)$).  But
what about specifying how the stars are distributed in binary or
higher-order multiple systems? These provide star-relevant internal
degrees of freedom that affect the stellar-dynamical and astrophysical
evolution of stellar populations.

Concerning the stellar IMF, the number of stars with masses in the
interval $m$ to $m+{\rm d}m$ born in one embedded cluster is
${\rm d}N=\xi(m)\,{\rm d}m$. Its dependence on the metal abundance and
density of the embedded cluster is reasonably well known (Kroupa {\it
  et al.}, 2024). The embedded clusters form in molecular cloud clumps
with a star-formation efficiency (SFE) near 30~per cent with half-mass
radii $r_{\rm h} < 1\,$pc (Marks \& Kroupa 2012; Zhou, Kroupa \& Dib,
2024) and masses in stars, $M_{\rm ecl}$, ranging from about a dozen
stars to many millions of stars, depending on the properties of the
hosting molecular clouds as set by the host-galaxy's state (e.g. in
self-regulated equilibrium or interacting).  Observations of nearby
molecular cloud clumps with ongoing star formation suggest the SFE to
lie between a~few and~30~per cent (e.g. Lada \& Lada, 2003, fig.~25 in
Megeath {\it et al.}, 2016, fig.~8 in Zhou, Dib \& Kroupa, 2024) while
Nbody calculations of the very well observed $1-2\,$Myr old Orion
Nebula Cluster (mass $\,\approx 10^3\,M_\odot$), the NGC3603 very
young cluster (mass $\,\approx 10^4\,M_\odot$), and the star-burst
cluster R136 (mass $\,\approx 10^5\,M_\odot$) suggest
SFE$\,\approx 33$~per cent in all cases (Kroupa, Aarseth \& Hurley,
2001; Banerjee \& Kroupa, 2018). A SFE near 33~per cent may be the
characteristic effective SFE involved in an embedded cluster emerging
from a molecular cloud clump.  The region-wide or galaxy-wide IMF
(gwIMF) follows by adding all stellar IMFs of all formed embedded
clusters in the region, leading to the gwIMF evolving with the metal
abundance and gas density (Jerabkova {\it et al.}, 2018).

Can a similar description of the binary-star properties of different
stellar populations or galaxies be arrived at? A problem facing this
goal is that no successful star-formation theory can be resorted to in
order to perform this synthesis, since at present the simulations of
star formation have neither achieved sufficient realism to match the
small embedded clusters in the nearest star-forming cloud in
Taurus-Auriga nor Orion Nebula type clusters (for a discussion see
Kroupa {\it et al.}  2024).

\section{Are stars born in  binary or in higher-order multiple stellar systems?}

Owing to the conservation of angular momentum pre-stellar molecular
cloud cores within a clump form not one star but multiple stellar
systems. These have stellar-dynamical decay time-scales of
$10^4-10^5\,$yr leaving, within this time-scale, a population
consisting of 50~or 33~per cent binaries and 50~or 66~per cent single
stars if all cores were to spawn non-hierarchical triple or quadruple
systems, respectively. Because $\approx 1\,$Myr old populations in
low-density star-forming regions have a binary fraction near~1 this
must imply that the vast majority of all stars form as binaries or
hierarchical, dynamically stable higher-order systems (Goodwin \&
Kroupa, 2005). The fraction of higher-order multiple systems is
sufficiently small (Goodwin {\it et al.}, 2007; Duch{\^e}ne \& Kraus,
2013; Offner {\it et al.}, 2023; Merle, 2024) to be neglected in
deriving the relevant distribution functions. The above is true for
late-type stars less massive than a few~$M_\odot$ but more massive
stars appear to be born with a triple and quadruple fraction near to
unity (Offner {\it et al.}, 2023). Thus, it may be assumed to
sufficient approximation that the total fraction of multiple systems
is $f_{\rm mult, birth} = 1$ at birth. Here
\begin{equation}
f_{\rm mult} = N_{\rm mult} / \left(N_{\rm sing}+N_{\rm mult}\right)  \, ,
\label{pk_eq:fbin}
\end{equation}
where $N_{\rm mult}$ is the number of multiple systems and
$N_{\rm sing}$ is the number of single stars in a star-count
survey. For example, if we survey 100 main-sequence systems on the sky
and 50 are binaries, 15 triples and 5 quadruples then
$f_{\rm mult, ms} = 0.7$, while the observer misses~95 stars if none
of the multiple systems are resolved. Triples and quadruples can thus
be counted as binaries, setting
$f_{\rm bin}\approx f_{\rm mult} = 0.7$ in this example.

\section{The distribution functions}

Given that late-type stars (massive stars obey different pairing
rules, see e.g. Dinnbier {\it et al.}, 2024; the birth
orbital-parameter distribution function defining massive stars still
needing to be formulated) appear to form as binaries, secondarily to
the stellar IMF, three additional birth distribution functions are
needed to specify the dynamical internal degrees of freedom of a just
born stellar population: the period distribution function,
$f_{\rm P, birth}(P)$, the mass-ratio distribution function,
$f_{\rm q, birth}(q)$ ($q=m_2/m_1\le 1$ with $m_1, m_2$ being the mass
of the primary and secondary, respectively), and the eccentricity
distribution function, $f_{\rm e, birth}(e)$.
${\rm d}N= f_{\rm x, birth}(x)\,{\rm d}x$ is the number -- or fraction
amongst all systems in a survey -- of binary systems with orbital
element $x$ in the range $x$ to $x+{\rm d}x$, with $x=P, q, e$. Note
that $P$ is customarily in days. If $a_{\rm AU}$ is the orbital
semi-major axis in astronomical units then by Kepler's third law,
$m_{\rm sys}=m_1+m_2=a_{\rm AU}^3/P_{\rm yr}^2$, where
$P_{\rm yr}=365.25\,P$ is the orbital period in years.

The following questions need to be answered: what are the mathematical
forms of $f_{\rm P, birth}, f_{\rm q, birth}, f_{\rm e, birth}$, and
if these exist, are these as dependent on the physical conditions of
the gas clouds as is $\xi(m)$ (see ``The Universality Hypothesis'' in
Kroupa 2011)?  Answering these questions is not trivial because none
of the distribution functions are observable: at any time a population
of stars is surveyed, it is already at an evolved astrophysical stage
(massive stars have evolved) and in an evolved stellar-dynamical state
because stars begin to gravitationally interact in the dense embedded
clusters as soon as they become ballistic particles. The properties of
the binary population thus changes on the crossing time scale of the
embedded cluster which is $\simless 0.3\,$Myr. The initial
distribution functions thus never exist physically, but they are
required for mathematically modelling stellar populations. They are
``hilfskonstrukts'' (Kroupa \& Jerabkova, 2018).

For example consider a very low-mass embedded cluster in which a
handfull of (say 4) binaries are observed to have formed within a
clump with a half-mass radius $r_{\rm h} \approx 0.1\,$pc. With an
average stellar mass of $0.3\,M_\odot$ and a star-formation efficiency
of $1/3$ the clump mass becomes $M_{\rm cl}\approx 8\,M_\odot$ such
that it's crossing time is
$t_{\rm cross} = 2\,r_{\rm h}/\sigma \approx 0.33\,$Myr, with
  \begin{equation}
    \sigma\approx \left(G\,M_{\rm cl}/r_{\rm h}\right)^{1/2} \, ,
\label{pk_eq:sigma}
  \end{equation}
  being the nominal velocity dispersion and $G$ the gravitational
  constant. Any wide binary that forms in such a low-density molecular
  cloud clump, i.e. a very low-mass embedded cluster, will thus have
  experienced a number of mild encounters with other binaries at an
  observation time of $0.5-1$Myr, with (in this case very mild)
  stellar-dynamical ejections likely having removed a few stars
  already by this time (Kroupa \& Bouvier, 2003). More massive
  embedded clusters are more stellar-dynamically active therewith
  dynamically processing the initial population on a shorter time
  scale (Kroupa, 1995a). 

{\it Important note}: In order to not affect the stellar IMF it is of
paramount importance, when constructing an initial stellar population,
to first create an array of stellar masses from a given $\xi(m)$ and
to then pair stars chosen from this array into binaries or
higher-order multiples such that some $f_{\rm q}$ is obeyed (Oh \&
Kroupa, 2016; Dinnbier {\it et al.} 2024; Dvorakova {\it et al.},
2024).

\begin{figure}
  \centerline{\includegraphics[width=0.85\textwidth,clip=]{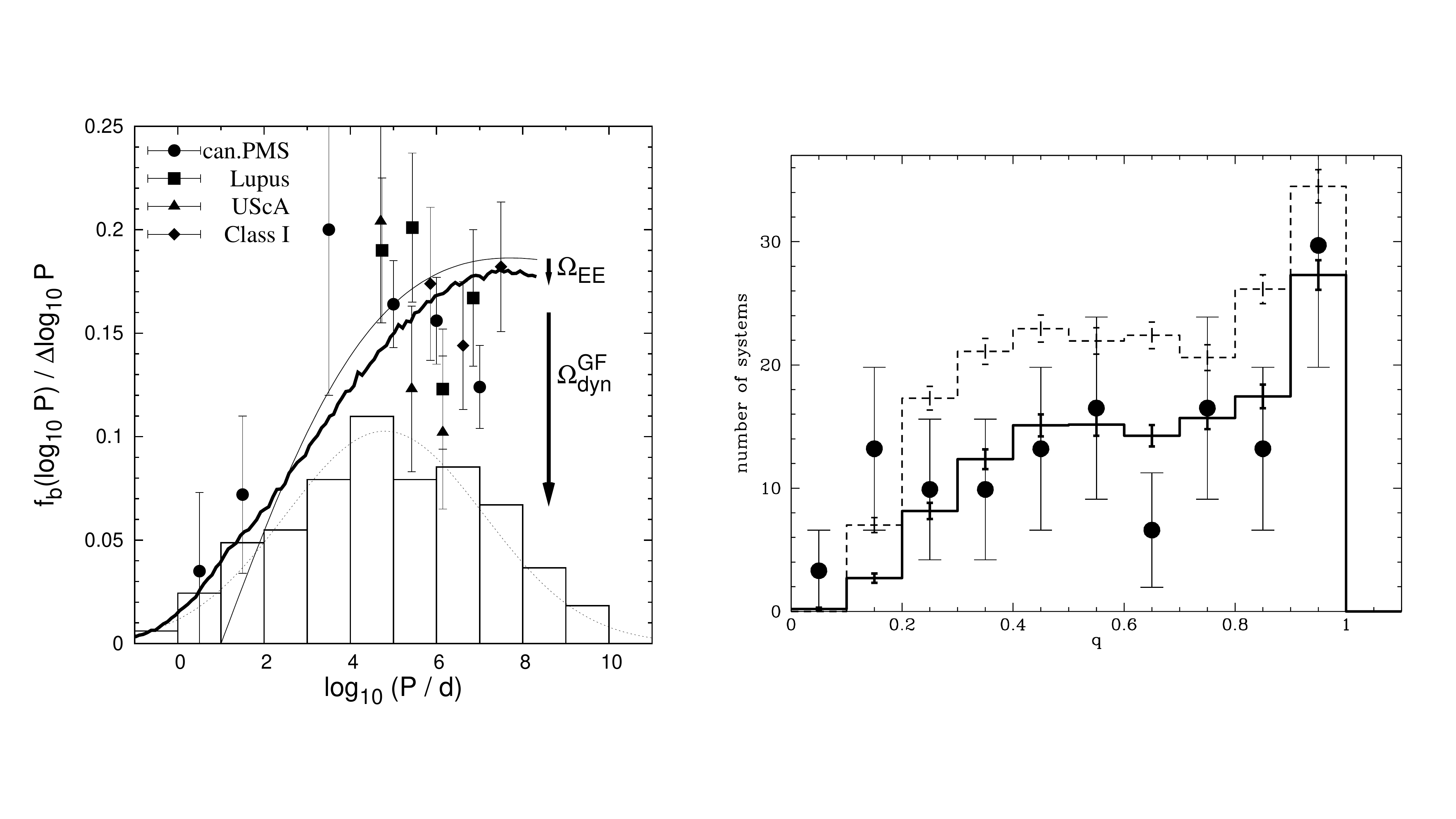}}
 \vspace{-12mm}
 \caption{{\it Left panel}: The distribution functions of periods,
   $f_{\rm b}=f_{\rm bin}$, of about $1\,$Myr old late-type binary
   systems in star-forming regions as shown by solid symbols. The
   histogram shows the period distribution of main sequence G-dwarfs
   in the Galactic field (Duquennoy \& Mayor 1991; Raghavan {\it et
     al.} 2010) which also matches that of K- and M-dwarfs that have a
   slightly lower binary fraction (see Fig.~\ref{pk_fig:fpmass}).  For
   example, 8~per cent of all field-G~dwarfs have a stellar companion
   with $10^3 < P/{\rm d} \le 10^4$.  The thin solid line is the birth
   distribution of periods, $f_{\rm P, birth}$
   (Eq.~\ref{pk_eq:initial}). It evolves through pre-main sequence
   eigenevolution ($\Omega_{\rm EE}$) in about $10^5\,$yr per binary
   to the thick solid curve which is the initial period distribution
   function, $f_{\rm P, init}$. The stellar-dynamical encounters in
   the embedded clusters in which the stellar population is born in
   evolve (by the stellar dynamical operator,
   $\Omega^{\rm GF}_{\rm dyn}$) $f_{\rm P, init}$ to the observed
   period distribution function, $f_{\rm P, ms}$, in the Galactic
   field shown as the thin-dotted curve. $\Omega^{\rm GF}_{\rm dyn}$
   is a function of the distribution of masses of the embedded
   clusters, $M_{\rm ecl}$, and their half-mass radii, $r_{\rm
     h}$. Adapted from fig.~1 in Marks, Kroupa \& Oh (2011), for
   further details see there.  {\it Right panel}: The mass-ratio
   distribution for primaries with masses
   $0.1 \le m_1/M_\odot \le 1.1$.  The solid circles show the mass
   ratio distribution of the volume-limited Solar-neighbourhood sample
   of observed stars (Reid \& Giziz, 1997). The dashed histogram is
   the initial ($f_{\rm q, init}$, Eq.~\ref{pk_eq:initial}) and the
   solid one is the model main-sequence Galactic-field population,
   $f_{\rm q, ms}$, computed according to Eq.~\ref{pk_eq:Omega}.  Note
   that the stellar dynamical processing in the birth embedded
   clusters reduces the number of binaries at all values of $q$ and
   how the pre-main sequence eigenevolution leads to a near-perfect
   match between the observed peak and the model near $q\approx 1$
   (the ``pre-main sequence eigenevolution'' peak in $q$).  Adapted
   from fig.~17 in Kroupa (2008), for details see there.  }
\label{pk_fig:lPandq}
\end{figure}

Fig.~\ref{pk_fig:lPandq} demonstrates the problem: About 1~Myr old
populations in star-forming regions of low density have a
significantly higher binary fraction than the main sequence (ms) stars
in the Galactic field which have an average age of about 5~Gyr. This
difference in the binary fractions, $f_{\rm bin, birth}\approx 1$ vs
$f_{\rm bin, ms}\approx 0.5$, is readily explained through the
disruption of binaries in the embedded clusters in which they formed
through stellar-dynamical encounters.

Before describing the procedure how the birth distribution functions
were arrived at a historical note is useful to appreciate the
developments: By the late 1980s it was already well known that (i)
binary systems form through three-body encounters in star clusters and
these slow down the Nbody computations significantly, especially if
multiple binary systems interact in the core of the clusters. And (ii)
surveys of field stars were showing many of them to be binary systems
(e.g. Abt 1983) such that some initial binary population appeared to
be necessary when initialising the Nbody models.  Therefore Sverre
Aarseth spent much effort developing integration, transformation and
regularisation methods to allow his Nbody codes to computationally
handle such multiple systems while the stars evolve
astrophysically. But the link between star clusters being the origin
of the field population was not significantly entertained (e.g. Wielen
1971).  In late-1992, a very excited Sverre Aarseth approached me at
coffee time at the Institute of Astronomy in Cambridge and truly
dragged me to his office to explain the newly-found implementation of
chain regularisation into the Nbody code (later published in Mikkola
\& Aarseth, 1993). I was not able to follow the technical details of
his explanations (being initially preoccupied with a cookie that had
disagreeably melted into my coffee cup and with merely counting stars
to obtain $\xi(m)$), but it dawned on me that this discovery improved
his Nbody codes dramatically, making the Nbody integration of star
clusters with a large initial binary population even more accessible
(his codes were the most advanced already with fully code-integrated
stellar evolution and the most advanced mathematical transformations
of the equations of motions for efficient and accurate time
integration).  This development allowed the code to efficiently
compute the dynamical interaction of compact sub-systems of evolving
stars, and it just so happened to occur when the first high-resolution
imaging and speckle interferometry observations of nearby star-forming
regions done in the early 1990s (see Kroupa 1995a for details) turned
up the unexpected result that nearly all pre-main sequence stars there
are binaries. I had witnessed these results when they were reported
for the first time at a conference on binary systems in Atlanta
in~1992, and it became apparent to me that the binaries probably break
up in embedded clusters. But this needed to be shown with Nbody
calculations. This is where the new chain regularisation came into
play.  The above demonstrates how different research developments can
conspire to interfere constructively at just the right time to allow
useful progress. The Aarseth suit of programmes (Aarseth 1999, 2003,
2008) contains a stand-alone fortran chain-regularisation code
(chain.f) for integrating small-$N$ systems, and given the importance
of chain regularisation, a newer C-code, CATENA, was later developed
by Jan Pflamm-Altenburg and applied to a Nbody study of the Trapezium
star cluster (Pflamm-Altenburg \& Kroupa 2006).

The procedure to derive the birth (and initial) distribution functions
is as follows: the Galactic field population needs to be reproduced
given the pre-main sequence constraints
(e.g. Fig.~\ref{pk_fig:lPandq}).  The constraint that
$f_{\rm bin, birth}\approx 1$ for the pre-main sequence population
stemming from the nearby (and thus common low-density) molecular cloud
clumps suggests the birth period distribution function to be rising
with $P$, since, by the Heggie-Hills law
(Sec.~\ref{pk_sec:physicalvariation}) the close binaries are not
significantly affected by stellar-dynamical encounters while the
long-period ones are likely to be disrupted. Indeed, the data in the
left panel of Fig.~\ref{pk_fig:lPandq} tell us that the binary
fraction for $P<1000\,$d ($f_{\rm bin}(P<10^3\,{\rm d})\approx 13$~per
cent) is similar in the pre-main sequence population as it is for the
metal-rich Galactic-field population in the Solar neighbourhood. In
order to arrive at a mathematical formulation of the plausible birth
distribution functions, Kroupa (1995a) applied an iterative
procedure. By performing Nbody simulations with an Aarseth (1999,
2003, 2008) code of star clusters until their dissolution (such that
their stars become field stars) with different birth densities and
starting with a first guess for
$f_{\rm P, birth}, f_{\rm q, birth}, f_{\rm e, birth}$, the
destruction of binary orbits through stellar-dynamical encounters was
quantified.  Sverre had to change the memory array structure in his
Nbody software so I could do the computations of star clusters
initially consisting of 100~per cent binary systems, a problem
considered to be sacrilegious and heretical until then (Aarseth, 1993,
private communication). For example, the seminal work on ``The
evolution of open clusters'' by the PhD student of Sverre Aarseth,
Elena Terlevich (1987), still relied on single-star simulations with
an earlier Aarseth Nbody code.  Returning to the present issue, a
second round of simulations used improved
$f_{\rm P, birth}, f_{\rm q, birth}, f_{\rm e, birth}$, with this
being repeated until these distribution functions were consistent with
the observational constraints evident in the left panel of
Fig.~\ref{pk_fig:lPandq} for the pre-main sequence stars and the field
stars.  ``Consistent'' here means according to inspection by eye. It
would be wrong to attempt a formal fit, because the pre-main sequence
observational data are for $\approx 1\,$Myr old systems that have
already been dynamically processed to some degree despite stemming
from low-density molecular clumps. And the Galactic field population
stems from the superposition of many dissolved embedded clusters, the
distribution of which is unknown. The choice of action was therefore
to seek $f_{\rm x, birth}$ such that an overall good reproduction of
the $\approx 1\,$Myr and $\approx 5\,$Gyr old populations could be
achieved through Nbody modelling.  An important cross check was to
ensure that the computed mass ratio and orbital eccentricity
distributions were also, at the same time, consistent with the data
available for the pre-main sequence and the field binaries. {\it This
  uniquely demonstrated that the field population can be arrived at by
  the stellar-dynamical processing of the birth population of binaries
  that is consistent with the observed binary systems in a pre-main
  sequence stellar population. For this to work, stars must form as
  binary systems in embedded clusters, implying that embedded clusters
  are the fundamental building blocks of galaxies (Kroupa 2005). }

The next step was to account for the observed correlations between
$P$, $q$ and $e$ for systems with $P\simless 10^3\,$d. These
correlations are obtained from the birth distributions via {\it
  pre-main sequence eigenevolution} which models the dissipative
evolution of these three internal degrees of freedom during the first
$\approx 10^5\,$yr of a binary's existence. To achieve a useful
mathematical description of this, theoretical work on the tidal
dissipation and tidal circularisation was accommodated (Kroupa,
1995b). Pre-main sequence eigenevolution leads to the shrinkage of an
orbit, to its circularisation and to a change in the mass ratio, $q$,
as the secondary component in the binary accretes gas from the
accretion disk of the primary. The changes in $P$, $q$ and $e$ occur
in dependence of the birth $P$.  The formulation of pre-main sequence
eigenevolution was improved slightly by Belloni {\it et al.} (2017)
who took into account the properties of stellar populations in
globular star clusters. Pre-main sequence eigenevolution thus changes
the masses of some of the stars and leads to mergers such that the
{\it initial} population (after the $\approx 10^5\,$yr evolution) has
a larger mass in stars by a few percent and contains a few~per cent
mergers of all initial stars which are now single stars.

The so-arrived at comprehensive and computationally highly efficient
approach was tested against the field population of nearby (and thus
fully resolved truly single) and more distant (and thus unresolved
binary) stars in terms of star counts, i.e. their respective stellar
luminosity functions, by computing star clusters until their
dissolution (Kroupa 1995c). An additional test of the above birth
distribution functions plus pre-main sequence eigenevolution theory
was performed against the observations of the density profile, the
velocity dispersion profile, the IMF and observed binary-star
properties in the Orion Nebula Cluster and the Pleiades cluster
(Kroupa, Aarseth \& Hurley 2001). These calculations also demonstrated
how an embedded cluster emerges through gas expulsion from its
molecular cloud clump to become part of an OB association and a
long-lived open star cluster.

In order to allow the rapid synthesis of entire stellar-field
populations in galaxies assuming all stars form as binaries and in
embedded clusters, the above was formalised in terms of operators that
act on $f_{\rm P, birth}, f_{\rm q, birth}, f_{\rm e, birth}$ (Kroupa
2002).  The mathematical formulation of the operators was achieved by
Marks, Kroupa \& Oh (2011) in terms of the {\it pre-main sequence
  eigenevolution operator}, $\Omega_{\rm EE}$ and the {\it
  stellar-dynamical operator}, $\Omega_{\rm dyn}$. After applying
$\Omega_{\rm EE}$ to evolve the birth to the initial distribution of
binaries in one computational step, the population is
stellar-dynamically processed by a second computational step by
applying $\Omega_{\rm dyn} = \Omega_{\rm dyn}(M_{\rm ecl}, r_{\rm h})$
to obtain the dynamically processed initial population after a
specified time. Because further dynamical processing of this remaining
binary population largely halts after gas expulsion expands the
cluster, freezing the stellar population and its binaries (apart from
individual hardening events), the frozen population in a
post-gas-expulsion young open cluster becomes its contribution to the
Galactic field population.

In summary, we can write that at a given $x$, the operator
$\Omega_{\rm dyn}$ reduces or increases $f_{\rm x}$ as a consequence
of the stellar-dynamical processing in the cluster characterised by
$M_{\rm ecl}, r_{\rm h}$, i.e. its density.  We thus have the
evolution of the birth population to the frozen main sequence (ms) one
being computed via the following steps:
\begin{equation}
  f_{\rm x, init}(x) = \Omega_{\rm EE} \, f_{\rm x, birth}(x) \;  \Longrightarrow
  \; f_{\rm x, ms}(x) = \Omega_{\rm dyn}(M_{\rm ecl}, r_{\rm h}) \,
    f_{\rm x, init}(x) \; . 
    \label{pk_eq:Omega}
  \end{equation}
  Thereafter the same step as in the first part of
  Eq.~\ref{pk_eq:Omega} can be performed using the main-sequence
  version of $\Omega_{\rm EE}$ in order to correct the orbital
  elements of the short-period ms binaries for tidal circularisation
  (Kroupa, 1995b).  The integration over (i.e. summation of) all such
  $\Omega_{\rm dyn}$-processed binary populations yields the field
  stellar population as it emanates from the population of embedded
  clusters that form at a given time. This is represented in
  Fig.~\ref{pk_fig:lPandq} by $\Omega^{\rm GF}_{\rm dyn}$.

  Therewith a formulation of the binary-star birth binary distribution
  functions that are valid for star formation in the Solar vicinity of
  the Galaxy is obtained,
\begin{align}
\xi(m) \quad & {\rm eq.~5-8  \;  in\; Kroupa \; {\it et \; al.} \;
               (2024)}, 
  \nonumber \\
f_{\rm P, birth}(q), f_{\rm q, birth}(P), f_{\rm e, birth}(e) \quad  & {\rm eq.~3b-8\; in \;
  Kroupa \; (1995b),} \; \nonumber \\ & {\rm updated \; in \; Belloni
                                       \; {\it et \; al.} \; (2017)}.
\label{pk_eq:initial}  
\end{align}

\section{Using the initial distribution functions}

The usage of the stellar IMF, $\xi(m)$, for various problems in
galactic, extra-galactic and high-redshift problems is well documented
(Hopkins 2018; Kroupa {\it et al.}, 2024).

Similarly, with the birth distribution functions,
$f_{\rm P, birth}(q), f_{\rm q, birth}(P), f_{\rm e, birth}(e)$, in
hand and assuming these to be invariant to the physical conditions,
(Carney {\it et al.} 2005; Kroupa 2011) any population of any age can
be synthesised.  For example, {\it forbidden binaries} must exist
(Kroupa, 1995b) -- these are binaries that have been hardened through
an encounter and appear above the eccentricity--period cutoff that
arises from tidal circularisation. Such forbidden binaries circularise
on the orbital time scale and allow insights into stellar interiors.

We can calculate the evolution of the birth binary population in the
Orion Nebula Cluster and the Pleiades (Kroupa, Aarseth \& Hurley,
2001).  The initial conditions under which the globular cluster
$\omega$~Cen, with its very small present-day binary fraction (Wragg
{\it et al.}, 2024) and its associated retrograde field population
(Carney {\it et al.}, 2005) most likely formed in can be inferred
(Marks, Kroupa \& Dabringhausen, 2022).  This system is interesting,
because Carney {\it et al.}  (2005) note that the prograde metal-poor
population of stars has an observed binary fraction that is not
distinguishable from that of the metal-rich population, while the
retrograde field population co-orbiting with $\omega$~Cen and of the
same low metallicity has a much smaller binary fraction. These
differences can be understood if the retrograde population was born in
one or more dense star-burst clusters, the remnants of which today are
evident as the ancient $\omega$~Cen.  The implied high birth density
of $\omega$~Cen of $\approx 10^8\, M_\odot/$pc$^3$ (see fig.~10 in
Marks {\it et al.}  2011) is of relevance also for the massive
globular cluster 47~Tuc which is reported to likewise have an
extremely low present-day binary fraction (M\"uller-Horn {\it et al.},
2024).

Further, the relevance of star-cluster-induced binary-star mergers for
the population of B[e] stars can be assessed: about 26--30~per cent of
all O-type stars, 13-24~per cent of all B-type stars and 5-8~per cent
of all A-type stars would be mergers (Dvo{\v{r}}{\'a}kov{\'a} {\it et
  al.}, 2024). A substantial fraction of massive stars are thus
probably mergers that were induced to happen through stellar-dynamical
encounters in their young star clusters where they were born. The
birth binary fraction of unity leads to profuse stellar-dynamical
ejections of~O and B~type stars from their birth embedded clusters as
shown by Nbody calculations (Oh \& Kroupa, 2016) and as is now
observed to be the case based on Gaia data (e.g. Stoop {\it et al.},
2024; Carretero-Castrillo {\it et al.}, 2024).

We can also synthesise galactic field populations by adding the many
contributions from all embedded clusters that have since dissolved
leading to the prediction that star-forming late-type dwarf galaxies
will have an overall binary fraction amongst its late-type stars of
$f_{\rm bin, ms}\approx 0.8$, Milky-Way-type galaxies have
$f_{\rm bin, ms}\approx 0.5$ (as is observed) and massive elliptical
galaxies have $f_{\rm bin, ms}\approx 0.35$ (Marks \& Kroupa, 2011).
The dynamical evolution of the Cepheid population in a galaxy can be
computed, giving insights, for example, into out-of-age Cepheids which
are like blue stragglers and form after a binary system merges into
the Cepheid instability strip (Dinnbier, Anderson \& Kroupa, 2024).

\begin{figure}
  \centerline{\includegraphics[width=0.65\textwidth,clip=]{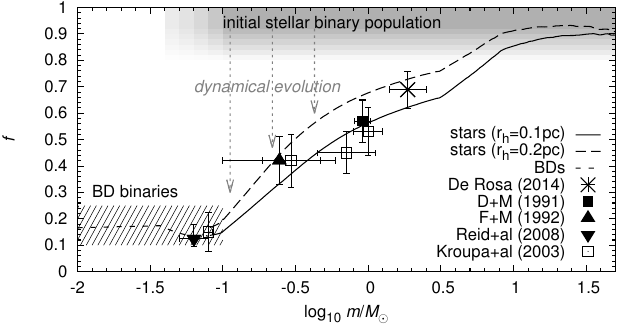}}
 \vspace{-4mm}
 \caption{The binary fraction as a function of the primary mass. Stars
   are born in embedded clusters with half mass radii $r_{\rm h}$ as
   binaries. This birth stellar binary population has
   $f=f_{\rm bin, birth}(m_1)\approx 1$ {\it for all primary masses},
   $m_1$. The stellar-dynamical encounters within the embedded
   clusters decrease $f_{\rm bin, birth}(m_1)$ more for smaller $m_1$
   because less-massive binary systems are, on average, less
   bound. The calculations (via $\Omega_{\rm dyn}$) show the decrease
   to match the observed data of old Galactic field populations
   (depicted by the different symbols) very well.  Brown dwarfs (BD)
   follow fundamentally different distribution functions to those of
   stars, and massive stars ($m_1> 5\,M_\odot$) likewise have
   different pairing rules but appear to be born typically in triple
   and quadruple systems. This figure demonstrates that it is
   incorrect to take the observed decrease of the binary fraction with
   decreasing primary-star mass as a direct constraint on
   star-formation theory. Adapted from fig.~6 in Thies {\it et al.}
   (2015), for details see there. }
\label{pk_fig:fpmass}
\end{figure}

Most of the above results relied heavily on the Aarseth Nbody codes to
do the time-forward integrations of the equations of motions of the
systems being studied. But these are CPU-time consuming and the
development of the $\Omega$-operators based on the Nbody results lead
to the BIPOS1 programme.  The programme BInary POpulation Synthesis~1
(BIPOS1) has been made publicly available to allow the rapid
computation of dynamically processed stellar populations
(Dabringhausen, Marks \& Kroupa, 2022). The modelled population of
galactic field stars turns out to be in excellent agreement with the
observed population of M-, K-~and G~dwarfs (see Marks \& Kroupa, 2011;
Cifuentes {\it et al.} 2024), accounts for the peak at $q\approx 1$
(Fig.~\ref{pk_fig:lPandq}) {\it after} the above theory was
formulated, and also reproduces the correlation between the binary
fraction and the mass of the primary star observed for Galactic field
stars (Fig.~\ref{pk_fig:fpmass}).

\section{The physical variation of the birth/initial distribution
  functions}
\label{pk_sec:physicalvariation}

The variation of $\xi(m)$ with metallicity, $Z$, and density of the
embedded cluster is reasonably well understood (Kroupa {\it et al.},
2024). From the empirical evidence it transpires that the
gravitationally unstable gas clump which forms an embedded cluster
does what theory broadly expected all along: in broad and simple
terms, at high (e.g. Solar) metallicity, the collapse cannot be as
deep because slightly overdense gas cools rapidly and so fragments
into typically less massive proto-stellar cores, while at low
metallicity, the collapse is deeper reaching larger densities before
individual proto-stars emerge thus leading to a more top-heavy stellar
IMF in the embedded cluster that forms in the molecular cloud clump.

The three birth binary star distribution functions,
$f_{\rm P, birth}, f_{\rm q, birth}, f_{\rm e, birth}$, most likely
vary in an according and a coordinated fashion. In extracting their
variation from the available data, one needs to take into account that
a metal-poor gas cloud will collapse to a denser state than a
metal-rich one such that the ensuing low-metallicity embedded cluster
will lead to a lower overall binary fraction {\it if} the birth
distribution functions were to be invariant, because more of the
binary systems are disrupted in a denser embedded cluster. Since {\it
  hard} (i.e. close, see below) binaries have too small a cross
section for significant stellar-dynamical processing, the number of
these does not change significantly and this is independent of
metallicity. {\it Soft} (i.e. wide, see below) low-Z binaries would
however be depleted compared to their counterparts at
$Z_\odot$. According to Eq.~\ref{pk_eq:fbin}, the binary fraction for
close binaries would thus be slightly smaller at low than at high
metallicity but would be quite insensitive to $Z$, while the binary
fraction of wide binaries would be smaller at low $Z$ than at
$Z_\odot$. The data by Lodieu {\it et al.} (2025) may be consistent
with this interpretation -- see below.

However, more likely $f_{\rm x, birth}({\rm low}\,Z)$ will differ from
$f_{\rm x, birth}({\rm high}\,Z)$ by the same broad physics: the
proto-stellar core collapses to a higher density before two
proto-stars emerge probably leading to a higher fraction of close
binaries at low metallicity, as discussed by Mazzola {\it et al.}
(2020). Longer-period binaries may appear just as in the high-$Z$ case
through random associations in the molecular cloud filaments of
protostars for late-type stars (see Kroupa {\it et al.}  2024 for a
discussion).  Since the molecular cloud core evolves to a denser state
at low $Z$, it is probable that also the wider binaries will end up
having shorter $P$ values than their equivalent high-$Z$ counterparts.

  The hard/soft boundary is of importance here: {\it hard binaries}
  have orbital velocities that are larger than the bulk velocity
  dispersion in the embedded cluster in which they are born, while
  {\it soft binaries} have slower orbital velocities. {\it There is no
    universal hard/soft boundary apart from that for an embedded
    cluster at a given time.} The hard/soft boundary is proportional
  to $r_{\rm h}^{-{1/2}}$ (Eq.~\ref{pk_eq:sigma}) such that this
  boundary is at a higher velocity for a low-$Z$ embedded cluster than
  for a high-$Z$ cluster of equal mass. The binary-star semimajor
  axes, $a$, are also probably systematically smaller at low-$Z$ than
  at high-$Z$, scaling proportionally to $r_{\rm h}$. But because the
  orbital velocity of a binary system is proportional to $a^{-1/2}$
  (eq.~142 in Kroupa 2008), the fraction of hard to soft binaries may
  remain the same in an embedded cluster of low metallicity compared
  to an equal-mass cluster at high metallicity.  However, it may be,
  by the Heggie-Hills law (``{\it hard binaries harden, soft binaries
    soften}'', Heggie, 1975; Hills, 1975), that the denser environment
  in the metal-poor embedded cluster hardens the tight binaries such
  that the fraction of {\it close} binaries ends up being larger in
  the metal-poor population than in the metal-rich one.

  The above discussion may be relevant for the discovery through large
  surveys by Moe, Kratter \& Badenes (2019) and Mazzola {\it et al.}
  (2020) that iron-poor populations have a higher fraction of {\it
    close} binaries than metal-rich populations. These authors also
  find the fraction of close binaries to vary with the alpha-element
  abundances. The variation of the observed populations with the metal
  abundance thus appears to be complicated. However, what matters is
  that the cooling function of the gas depends primarily on the
  available atomic transitions such that $Z$ is the relevant
  quantity. Thus, a population that is alpha-element enriched but iron
  poor might have the same binary star properties to a population that
  is iron rich and alpha-element poor. For example, according to
  Forbes {\it et al.} (2011, see also Wirth {\it et al.}, 2022),
  $[Z/H]=[Fe/H]+[\alpha/H]$. Returning to the samples by Moe {\it et
    al.}  (2019, their fig.~3) and Mazzola {\it et al.} (2020, their
  fig.~5), their data appear to indicate that at near Solar
  metallicity ([Fe/H]$\,\approx 0$) the fraction of close
  ($P<10^4\,$d) binaries in a survey amounts to about 20~per cent
  which is consistent with Fig.~\ref{pk_fig:lPandq} (left
  panel). Fig.~3 in Moe {\it et al.}  (2019) suggests that at
  [Fe/H]$\,\approx -3$ this fraction increases to 60~per cent. In
  Fig.~6 in Mazzola {\it et al.} (2020) the published fraction reaches
  $>100\,$per cent at moderate low metallicity
  ([Z/H]$\,\approx -0.8$), leaving no room for a contribution to the
  stellar population by wider binaries (with $P>10^4\,$d).  In
  contrast to the above studies, Lodieu {\it et al.} (2025) find
  Solar-type stars at low ``metallicity'' ($[Fe/H]<-1.5$) to have a
  comparable/much smaller binary fraction to those at Solar
  metallicity for close binaries ($<8\,$AU)/wide binaries
  ($8-10^4\,$AU), respectively.  In view of the above results that
  appear to differ substantially in terms of the reported binary
  fraction of close binaries at low~Z, it is therefore unclear how
  these quoted fractions are to be compared and care needs to be taken
  in precisely defining what is counted how.

  The above information provides a suggestion how the current
  formulation of $f_{\rm x, \, birth/init}(Z\approx Z_\odot)$ given by
  Eq.~\ref{pk_eq:initial} can be generalised to other values of
  $Z$. An iterative process of matching the observational constraints
  (e.g. by Carney {\it et al.}, 2005; Moe {\it et al.}, 2019; Mazzola
  {\it et al.}, 2020; Lodieu {\it et al.}, 2025) to Nbody-generated
  data, similarly to that performed by Kroupa (1995a,b), will be
  necessary to infer a viable model. Hereby smaller $r_{\rm h}$ values
  at low $Z$ will need to be probed.  The so-arrived at model for
  $f_{\rm x, \, birth/init}(Z)$ will then be of importance to test and
  check the theory of star formation (rather than the other way
  around).

In performing such tests of star-formation theory, it needs to be
remembered that computer simulations of collapsing molecular cloud
cores do not produce a measurable $f_{\rm x, \, birth/init}(Z)$ for
the same reasons why these functions cannot be measured on the sky
(the forming binaries break up due to encounters in these simulations
while others are still forming).  The theoretically produced
$f_{\rm x, \, birth/init}(Z)$ can however be extracted by tracing the
individual stellar particles back to where and in which stellar system
they were born in, in order to quantify as well as is possible the
values of $P, q, e$ for this system. The theoretical
$f_{\rm x, \, birth/init}(Z)$ can be constructed from the
distributions of the so obtained quantities.

Metallicity and density dependent operators
$\Omega_{\rm EE}(Z), \Omega_{\rm dyn}(r_{\rm h}, M_{\rm ecl}:Z)$ will
then need to be quantified along the lines as performed by Marks,
Kroupa \& Oh (2011) to allow rapid dynamical synthesising of stellar
populations being born in galaxies of different properties and at
different redshifts as probed by Marks \& Kroupa (2011) for the
$Z\approx Z_\odot$ case. 

\section{Conclusions}

The difference between the binary populations in star forming
  regions and the Galactic field can be  understood if
  stars form mostly as binary systems and in embedded star clusters in
  molecular cloud clumps. Dynamical processing in these embedded
  clusters dissolves a large fraction of the soft binaries. The binary
  properties of Galactic field stellar populations are therewith
  reasonably well understood at Solar metallicity.

  The Sun too may have had a binary companion orbiting it with a birth
  period $P\simgreat 10^6\,$d ($a\simgreat 300\,$AU) in order to be
  consistent with the current orbit of Neptune. Such orbits in fact
  dominate the birth period distribution function
  (Fig.~\ref{pk_fig:lPandq}).  The existence of planetary systems
  around most stars is thus not in contradiction to all stars forming
  as binary systems. The finding that many exoplanetary systems appear
  to be ``jumbled-up'', i.e. not as regular and orderly as our Solar
  System (e.g. Zhu \& Dong, 2021), may be due to the perturbations
  during the formation stage or thereafter by the not so distant
  companion star and stellar-dynamical encounters in their birth
  embedded clusters (Thies {\it et al.} 2011).

  There is evidence that low-metallicity populations
  follow the trend already well documented by the variation of the
  stellar IMF. The sense may be that with decreasing metallicity the
  average stars appear to be more massive and consist of closer
  binaries at birth, and probably also with more similar companion
  masses. There is thus emerging evidence how the three birth
  distribution functions,
  $f_{\rm P, birth}(P:Z, m_1), f_{\rm q, birth}(q:Z, m_1), f_{\rm e,
    birth}(e:Z, m_1)$, differ form their metal-rich
  ($Z\approx Z_\odot$) counterparts such that, upon the dynamical
  processing in the low-metallicity embedded clusters, the survey
  results are obtained. However, as discussed above, some of
  the evidence appears to be contradictory. 

  In terms of challenges for the future to improve the mathematical
  description of birth distribution functions of stars: (A)~the
  variation of these with metallicity and density of the embedded
  clusters in which stars are born will need to be studied. This will
  require Nbody simulations to quantify the stellar-dynamical
  processing of the distribution functions for plausible mass
  functions of embedded clusters that are likely to have spawned the
  stellar populations under study. For example, a field population
  might stem from one single massive star burst cluster or from a
  large population of less-massive embedded clusters. An example of
  this is the work on $\omega$~Cen and its associated field population
  mentioned in the text. (B)~The extension of the birth distribution
  functions to include triple and quadruple systems is needed. This is
  an interesting task to be mastered by seeking a mathematical
  description that preserves the above (Eq.~\ref{pk_eq:initial}) but
  accommodates additional companions without violating the stellar
  IMF.

  As a final comment, the above and associated work shows that there
  is no single correction for unresolved binary systems for a given
  star-count survey because the dynamical histories of observed
  populations can differ such that their binary properties differ. For
  example, a low-mass molecular cloud in which low-mass embedded
  clusters emerge spawns a T~Tauri association with a high binary
  fraction ($f_{\rm bin} \approx 0.8$). A neighbouring massive
  molecular cloud in which ONC- (or even NGC3603-) type star-burst
  embedded clusters condense spawns an OB~association with an overall
  binary fraction $f_{\rm bin}\approx 0.4-0.5$.  And, even if they are
  merely ``hilfskonstrukts'', the inference of the true birth/initial
  distribution functions are needed for efficiently modelling stellar
  populations. Finally, given some observational survey data,
  understanding the underlying stellar populations and their binary
  properties non-negotiably requires full understanding of their
  stellar-dynamical and astrophysical evolution. Without the
  monumentally fundamental Nbody-development work by Sverre Aarseth in
  Cambridge since the 1960s this would not have been possible to be
  achieved.

\acknowledgements This work would have been entirely impossible
without Sverre Aarseth (1934–2024), whose pioneering contributions to
Nbody coding and simulations profoundly shaped the field. His
mentorship and guidance were pivotal to my development as both a
researcher and a person as well as for the work reported
here. Sverre’s legacy as a brilliant scientist and a generous mentor
will continue to inspire the Nbody community. I sincerely thank Sverre
Aarseth for his immensely kind reactions to my computational needs,
incorporating the treatment of all stars as initial binaries, gas
expulsion and the inclusion of the computation of a cluster's tidal
tails. All these innovations to the codes allowed a significant amount
of research I was involved with, and I will be forever indebted to
Sverre for making this possible.  I also acknowledge the DAAD
Eastern-European Bonn-Prague exchange programme for support.
   

\end{document}